\documentstyle[prb,aps,eqsecnum]{revtex}
\begin{document}
\title
{Quantum antiferromagnet at finite temperature: a gauge field approach}
\author{Oleg A. Starykh \cite{oas}, George Reiter}
\address
{Texas Center for Superconductivity, University of Houston, Houston,
TX 77204-5932}

\maketitle
\begin{abstract}
Starting from the $CP^{N-1}$ model description of the thermally disordered
phase
of the $D=2$ quantum antiferromagnet, we examine the interaction of the
Schwinger-boson spin-$1/2$ mean-field excitations with the generated gauge
(chirality) fluctuations in the framework of the $1/N$ expansion.
This interaction dramatically supresses the one-particle motion,
but enhances the staggered static susceptibility. This means that
actual excitations in the system are represented by the collective
spin-$1$ excitations , whereas one-particle excitations disappear from
the problem. We also show that massive fluctuations of the constraint field
are significant for the susceptibility calculations. A connection
with the problem of a particle in random magnetic field is discussed.

\end{abstract}



Some time ago, two self-consistent approaches to the finite temperature
properties of the low-dimensional magnets were developed: the modified spin
wave theory [\onlinecite{takahashi}] and the Schwinger-boson theory
[\onlinecite{arovas}].
Both of them correctly describe the generation of the correlaton length by
the thermal fluctuations, by complying with the Mermin-Wagner theorem
[\onlinecite{mermin}].
 Due to
its conservation of the spin rotational invariance in the thermally (or
quantum)
 disordered (paramagnetic) phase the Schwinger-boson theory is a suitable
starting point. Moreover, it allows the
generalization of the Heisenberg Hamiltonian to that of an $SU(N)$-invariant
model [\onlinecite{arovas}], for which a powerful $1/N$ expansion can
be developed [\onlinecite{arovas,readsachdev}]. Perturbation expansion in a
parameter
$1/N$ around the mean-field ($N=\infty$) saddle point provides a convinient
way of treating the fluctuations in the isotropic system.

The applicability of the Schwinger-boson transformation of the spin operator
relies completely on the necessary constraint at every site: the
number of boson degrees of freedom has to be equal to $NS$, where $S$ is the
the value and $N$ is the number of boson "colors" [\onlinecite{arovas}]. Due to
the
constraint only $N-1$ out of $N$ bosons per site are really independent.
Usually
, this constraint is introduced through the Lagrange field $\lambda$, defined
on every lattice site. In the mean-field approximation on-site constraint
is approximated by the averaged one [\onlinecite{arovas}], and it makes all $N$
bosons
free.

As was shown in [\onlinecite{readsachdev}], the fluctuations of the constraint
field $\lambda$ above the mean-field value $m^2$ and the
phases of the Hubbard-Stratonovich decoupling field form three components of
the $U(1)$ gauge field in the continuum limit. Also it was proved that the
continuum limit
of the Schwinger-boson large-$N$ theory coincides with the same limit of the
$CP^{N-1}$ model at
distances much larger than the lattice spacing [\onlinecite{readsachdev}].

Despite the fact that $CP^{N-1}$ model at $N=2$ is just another representation
of the $O(3)$ nonlinear $\sigma$ model, there is no direct correspondence
between these models at arbitrary $N$. In principle, identical results should
be obtained from the exact solutions of both models for the case of physical
spins. It is the perturbative expansion around different $N=\infty$ saddle
points of the models that makes the predictions of the models very
different. Namely, mean-field analysis of the $CP^{N-1}$ model in the
disordered
 phase
produces two branches of elementary excitations with spin-$1/2$, whereas the
$O(N)$ nonlinear $\sigma$ model gives triply degenerate spin-$1$ excitations
in the disordered phase at $N=\infty$ [\onlinecite{yale}]. At the same time
expressions for the
spin-correlation length obtained from these models in the renormalized
classical
region are similar at the mean-field level.

In this communication we investigate the corrections beyond the mean-field
level in the $CP^{N-1}$ model for
the special case of the thermally
disordered phase (or renormalized classical phase at $T\ne 0$, which
corresponds
to the Neel-type state at zero temperature)
of the $CP^{N-1}$ model, by exploiting the abovementioned
equivalence of it to the continuum limit of the Schwinger-boson theory.
It is neseccary to stress here that in the analysis to follow we consider the
lowest-energy excitations with momentum (and energy) much smaller than the
inverse correlation length. In that regime thermal fluctuations of locally
Neel ordered regions destroy completely long range Neel order and
antiferromagnet is disordered, while inside the regions of the size of
correlation length
the system behaves as in the broken symmetry phase [\onlinecite{yale}].
 The model contains two collective
modes: the massless $U(1)$ gauge field (due to the famous mechanism of the
dynamical generation of the kinetic energy of the gauge field
 [\onlinecite{dadda,polyakov}])
and the massive fluctuations of the
constraint field $\lambda$.
$\lambda$-fluctuations produce only the short-range density-density
interaction term, whereas the coupling to the gauge field produces the
long-range current-current interaction. Due to their masslessness, the gauge
fluctuations change significantly the mean-field picture. As we will show here,
it is the interaction with the generated gauge field that supresses spin-$1/2$
excitations, while only slightly affecting collective excitations with
spin $1$.
At the same time, interaction of the massive $\lambda$-fluctuations with the
mean-field excitations is much weaker, nevertheless it gives a
susceptibility correction of the same order of magnitude as the interaction
with gauge fluctuations.

Effect of the interaction with generated gauge field on the properties of
free spin-$1/2$ spinons was investigated in ref.[\onlinecite{readsachdev}] for
the
case of quantum disordered phase of the $SU(N)$ quantum antiferromagnet, and it
was found that due to the "hedgehog"-like instanton tunneling events spinons
are confined into the spin-$1$ pairs. Contrast to these we are considering
the thermally disordered phase without topological defects.

Our starting point is the nonlinear $\sigma$ model Euclidian action in the
continuum limit
$$ {\cal S}=\frac{c}{2g}\int_{0}^{\beta} d{\tau} \int d^2 {\bf r} \left\{\frac
{1} {c^2}(\partial_{\tau}{\bf n})^2 + (\partial_i{\bf n})^2\right\}, i=x,y
\eqno(1)$$
with the constraint ${\bf n}^2=1$. Vector ${\bf n}$ describes local staggered
magnetization. Putting $c=1$, using the $CP^1$ representation
${\bf n}=z^+{\bf \sigma}z$ ($z$ is a complex two-component field),
and introducing ${\bf A}_{\mu}=-\frac{i}{2}(z^+\partial_{\mu}z - (\partial_
{\mu}z^+)z)$  the action becames [\onlinecite{polyakov}]
$${\cal S}=\frac{1}{g}\int_{0}^{\beta} d{\tau} \int d^2 {\bf r}
\left\{|(\partial_{\mu} -
i{\bf A}_{\mu})z|^2 + i{\lambda}(|z|^2-1)\right\},  {\mu}= {\tau}, x, y
\eqno(2)$$
where Lagrange multiplier $\lambda$ enforces the constraint $z^+z=1$.

Generalizing the doublet $z$ to the $N$-component vector we arrive at the
$CP^{N-1}$
model action. The saddle point equation for the $\lambda$-field produces the
gap $m$
in the spectrum of $z$-quanta, and finally we get (we have rescaled $z$ by
$\sqrt g$)
$${\cal Z}=\int D z \int D {\bf A} \int D {\lambda'} exp-\int_{0}^{\beta}
d{\tau} \int d^2 {\bf r} \left\{
|(\partial_{\mu} - i{\bf A}_{\mu})z|^2 + m^2 |z|^2 + i{\lambda'} |z|^2
\right\},   \eqno(3)$$
where we have used parametrization [\onlinecite{polyakov,readsachdev}]:
$$i{\lambda}(x,\tau)=m^2 + i{\lambda'}(x,\tau),$$ with the condition
 $$\int_{0}^{\beta} d{\tau} \int d^2 {\bf r} {\lambda'}(x,{\tau})=0.$$

It is important for us that in the thermally disordered phase the gap $m$
is exponentially small as a function of temperature [\onlinecite{arovas}]
$$ m \simeq Texp-\frac{const}{T},       \eqno(4)$$

 We have briefly described the derivation of the $CP^{N-1}$ model from the
nonlinear $\sigma$ model for the sake of completeness of our discussion.
Microscopical derivation of the expression $(3)$ was done in
[\onlinecite{readsachdev}], where
the coupling of $\bf A$ field to the spin-Peierls order parameter was also
found
, but we will not consider this coupling here.

Integration over $z$ fields produces the effective action for ${\bf A}$ and
  ${\lambda'}$ fields.
In the large-$N$ limit, the gaussian (quadratic) fluctuations around
the saddle point $<{\bf A}>=0$, $<{\lambda'}>=0$ are leading,
because the higher orders of ${\bf A}$ (${\lambda'}$) will have additional
$\frac{1}{N}$ factors [\onlinecite{polyakov}]. This quadratic approximation is
equivalent to the
calculation of the polarization operator of $z$ fields. The corresponding
diagrams, describing generation of the gauge field energy [\onlinecite{dadda}],
are shown on Fig.1 and in the Coulomb gauge $(A_0=0)$ an explicit expression
for
the polarization is
$$ \Pi_{\mu \nu}(q,\omega_n)=-\frac{1}{\beta} \sum_{l} \int \frac{d^2 {\bf p}}
{(2\pi)^2} G(p,\epsilon_l) G(p+q,\epsilon_l + \omega_n) (2{\bf p}+{\bf q})_{\mu
} (2{\bf p}+{\bf q})_{\nu} - 2\delta_{\mu \nu} \frac{1}{\beta} \sum_{l}
\int \frac{d^2 {\bf p}}{(2\pi)^2} G(p,\epsilon_l) , \eqno (5)$$
where $G(p,\epsilon)$ is the Green's function of $z$-bosons in the absense of
gauge fields
$$ G(p,\epsilon_l)=-\frac{1}{{\epsilon_l}^2+{\bf p}^2 +m^2}$$
and $\epsilon_l=2\pi lT, \omega_n=2\pi nT$ are bosonic Matsubara frequencies.

Both one-loop integrals in $(5)$ are ultraviolet divergent but the
Pauli-Villars
regularization makes them finite. Because the gap $m$ is much smaller than the
temperature $T$, the lowest excitations (i.e. those inside the gap) correspond
to $\omega_n=0$ in $(5)$, which is equivalent to a static approximation. Making
an expansion over $\frac{{\bf q}^2}{m^2}$ in the integrand we get
$$ \Pi_{\mu \nu}(q,\omega_n=0)=\frac{1}{12\pi}(\frac{1}{\beta} \sum_{l} \frac
{1}{{\epsilon_l}^2 + m^2})(\delta_{\mu \nu}- \frac{q_{\mu} q_{\nu}}{{\bf q}^2})
{\bf q}^2. \eqno (6) $$
Using
$$ \frac{1}{\beta} \sum_{l} \frac{1}{{\epsilon_l}^2 + m^2}=\frac{1}{m}(
\frac{1}{2} + \frac{1}{e^{\beta m} -1}) \simeq \frac{1}{\beta m^2}, $$
we obtain the following expression for the Green's function of the gauge field
$$ D_{\mu \nu}(q)=-<T A_{\mu}(q)A_{\nu}(-q)>=-\frac{1}{N}\Pi_{\mu \nu}^{-1}(q)=
-\frac{12 \pi \beta m^2}{N {\bf q}^2} (\delta_{\mu \nu}-\frac{q_{\mu} q_{\nu}}
{{\bf q}^2}) , \eqno (7)$$
which describes static random magnetic fields with the correlator
$$<h(q)h(-q)> = \frac{12 \pi \beta m^2}{N}  \eqno (8)$$
Physically, the magnetic field h(r) describes the chirality through
[\onlinecite{polyakov}]
$$\frac{1}{2} {\bf n}{\cdot}(\partial_{\mu} {\bf n} \times \partial_{\nu} {\bf
n})=
\partial_{\mu} A_{\nu} - \partial_{\nu} A_{\mu} = h . \eqno (9)$$

The nature of static approximation is easy to understand: the minimal energy
of the $z$-boson is $m$, whereas ${\bf A}$ fields are lying inside the gap $m$
and, hence, they can be considered as the static ones in comparison with fast
$z$-quantas.

Knowing the Green's function, it is easy to calculate the contribution
of the gauge fluctuations to the specific heat
$$\delta C_v \simeq m^2 logT .$$
Because of the $m^2$ prefactor, this contribution is small.

The problem of a particle subject to random magnetic field has arisen in the
gauge theory of the $t-J$ model [\onlinecite{lee,ioffe}], in the vortex lines
dynamics of high-$T_c$ materials, and in the study of the transport properties
of the normal
film
-superconducting film sandwiches in a magnetic field [\onlinecite{sandvich}].
Significant progress has been made recently in understanding of the
problem, mainly due to the development of nonperturbative approaches
 [\onlinecite{wheatley}]. Due to the long range nature of the vector potential
${\bf A}$
and two-dimensionality of the problem, the interaction between the $z$-particle
and the gauge
field is
strong enough to produce divergences in the leading corrections to the
particle's
self-energy and vertex [\onlinecite{lee,ioffe}].

As we will show later, $z$-bosons also have similar divergences. The divergence
of
the self-energy implies that the averaged over the gauge fluctuations Green's
function is zero, or the $z$-particle does not propagate. But it is necessary
to keep in mind that
 $z$-boson describes an $S=\frac{1}{2}$ excitation at $N=2$, whereas physical
excitations (i.e. those corresponding to poles in the susceptibility) are
described
by the pair of the $z$-particles , which form an $S=1$ excitation
 [\onlinecite{chubukov}].

At the mean field level the susceptibility is given by the bubble diagram
(Fig.2), for an $N$-component boson system
$$ \chi^{mf}(q=0,\omega=0)=N \frac{1}{4 \pi \beta} \sum_{l} \frac{1}{
{\epsilon_l}^2 + m^2} = \frac{N}{4 \pi \beta m^2} . \eqno (10)$$
$\chi^{mf}(q=0,\omega =0)$ diverges at $T=0$ $(m=0)$ indicating the transition
into ordered Neel-type state
(since ${\bf n}$ is local $staggered$ magnetization, $q=0$ actually corresponds
to antiferromagnetic vector $(\pi,\pi)$ in the original problem).
The first $1/N$ correction to $\chi^{mf}$ is
generated by the self-energy insertions to the boson lines and by the vertex
correction
(Fig.3, solid line denotes $z$ propagator, wiggly - gauge-field propagator).
The corresponding analytical expressions are:
$$ \chi_a^{1/N}(q=0,\omega =0)=\frac{1}{{\beta}^2} \sum_{l} \int \frac{d^2
{\bf k}}{(2 {\pi})^2} \int \frac{d^2 {\bf q}}{(2{\pi})^2} G^3(k,\epsilon_l)
G(k-q,\epsilon_l)D_{\mu \nu}(q)(2{\bf k}-{\bf q})_{\mu} (2{\bf k}-{\bf q})_{
\nu}$$
$$ \chi_b^{1/N}(q=0,\omega =0)=\frac{1}{{\beta}^2} \sum_{l} \int \frac{d^2
{\bf k}}{(2 {\pi})^2} \int \frac{d^2 {\bf q}}{(2{\pi})^2} G^3(k,\epsilon_l)
D_{\mu \mu}(q) , \eqno (11)$$
$$ \chi_c^{1/N}(q=0,\omega =0)=\frac{1}{{\beta}^2} \sum_{l} \int \frac{d^2
{\bf k}}{(2 {\pi})^2} \int \frac{d^2 {\bf q}}{(2{\pi})^2} G^2(k,\epsilon_l)
G^2(k-q,\epsilon_l)D_{\mu \nu}(q)(2{\bf k}-{\bf q})_{\mu} (2{\bf k}-{\bf q})_{
\nu}$$

Diagram $3b$ can be calculated exactly, while expansion
over ${{\bf q}^2}$ is necessary in other expressions. Straightforward
calculations give the
following answer
$$\chi_i^{1/N}(q=0,\omega =0)=\frac{-1}{(2 \pi \beta)^2}\left\{\sum_{l}
\frac{1}{({\epsilon}_{l}^2 +m^2)^2} C_i \int \frac{d q}{q} + \sum_{l}
\frac{1}{({\epsilon}_{l}^2 +m^2)^3} S_i \int d q{ q} \right\}  \eqno (12)$$
Numerical coefficients $C_i$ and $S_i$ are collected in Table I, as well as
 symmetry factors ($F$) of diagrams.

As one should expect from the gauge invariance, the logarithmically divergent
term disappears from final expression for the first $1/N$
correction to susceptibility and, using $m$ as an upper cut-off in the
integration over $q$, we get (subscript $A$ denotes the gauge field origin of
this correction)
$$ \chi_{A}^{1/N}(q=0,\omega =0)= \frac{0.7}{4 \pi \beta m^2}  \eqno (13) $$

Unexpectedly, the susceptibility correction $\chi_{A}^{1/N}$ is positive
($\chi=\chi^{mf} + \chi_{A}^{1/N}$), which implies that
the gauge fluctuations enhance the short range antiferromagnetic order. This
result is not as strange as it at first appears. Let us recall the phenomenon
of the quantum stabilization of the classical order in the $J_1-J_2$ model
 [\onlinecite{frust}]:
quantum fluctuations act to reinforce the classical order beyond the
classical domain of stability.

The finite result for $\chi_{A}^{1/N}$ means that despite the dramatic
supression of the one-particle motion (i.e. the divergency of the
single-particle self-energy $-$ diagrams $3a$ and $3b$), the coherent
propagation
of a pair of $z$-particles is not affected significantly by the gauge
fluctuations. This, in turn, means that the final physical picture is
similar to that of the $O(N)$ nonlinear $\sigma$ model at $N=\infty$ : the
elementary excitations in the disordered phase of the antiferromagnet
are $spin-1$ excitations.

It is possible to show the origin of the obtained result using an eikonal
expansion [\onlinecite{fradkin}]. Analogous to [\onlinecite{lee}], the
self-energy divergence corresponds to the vanishing of the $z$-particle
Green's function. In the eikonal approximation, the Green's function depends
upon the field
${\bf A}$ through the phase $exp(i{\Gamma})=exp(i \int_{0}^{t} d{\tau} A_{\mu}
({\bf r}) \frac{d r_{\mu}}{d {\tau}})$, where ${\bf r}({\tau})$ is a straight
line path connecting the initial (${\bf r}(0)=0$) and final (${\bf r}(t)={\bf
r'}$) points. In the simplest approximation ${\bf r}({\tau})={\bf
r'}\frac{{\tau}}{t}$, hence
${\Gamma}=\int \frac{d {\bf q}}{(2\pi)^2} \int_{0}^{t} d{\tau} \frac{r'_{\mu}}
{t} A_{\mu}(q) e^{i{\bf q}{\bf r'} \frac{{\tau}}{t}} \simeq r'_{\mu}
 \int \frac{d {\bf q}}{(2\pi)^2} A_{\mu}(q)$ for small enough ${\bf q}$.
After averaging over gauge field fluctuations we have
$$ <e^{i{\Gamma}}>=exp-\frac{1}{2}<{\Gamma}^2> \simeq exp-\frac{1}{2}
r'_{\mu} r'_{\nu} \int \frac{d {\bf q}}{(2\pi)^2} <A_{\mu}(q)A_{\nu}(-q)>$$
Using eq.$(7)$ we arrive to the logarithmically divergent expression
$$<e^{i{\Gamma}}>=
 exp -\frac{1}{2} {\bf r'}^2 \frac{12\pi \beta m^2}{N} \int_{0}^{2\pi}
d{\theta} sin^2{\theta} \int_{0}^{m} \frac{dq q}{(2\pi)^2} \frac{1}{q^2}$$
Thus, $<e^{i{\Gamma}}>=0$ for any nonzero ${\bf r'}$. At the same time, in the
susceptibility calculations we are dealing with $closed$ trajectories
[\onlinecite{lee}], where by the Stokes theorem $\int d{\tau} {\bf A}({\bf
r})\frac{d{\bf r}}{d {\tau}}$ is the magnetic flux through the loop
formed by trajectory ${\bf r}({\tau})$ [\onlinecite{wheatley}]. According to
$(8)$ the
magnetic field correlator is constant. Therefore the momentum integration will
not
produce infrared divergency.

We also tried to calculate the $1/N$ correction to the staggered dynamical
susceptibility. In that case, the analogous cancelation of the logarithmically
divergent terms was found. As before, it is nothing but the consequence of the
gauge invariance of $\chi$. Unfortunately, no damping was found
for the frequency less than the gap. Thus, interaction with gauge field can
not eliminate (at least at $1/N$ order)
another shortcoming of the mean-field Schwinger-boson theory - the absence
of the relaxational modes in the system. Such hydrodynamical excitations
were found by the summation of the ladder diagrams describing the boson-boson
interaction beyond the mean-field approximation of the Schwinger-boson theory
 [\onlinecite{chubukov}].
In addition, recent study of the $O(N)$ quantum nonlinear
$\sigma$ model by an $1/N$ expansion [\onlinecite{yale}] shows that an
interaction
with massive fluctuations of the constraint field does produce damping
in the susceptibility and nonzero density of states inside the gap region.

Because of all these facts, we understand the problem as following:
the main role of the gauge fluctuations is not only to eliminate the mean-field
excitations from the spectrum, but to conserve the coherent pair propagation.
After that, the problem is equivalent to the initial nonlinear $\sigma$
model, which naturally describes $S=1$ excitations.

To confirm this point we will calculate the effect of the massive $\lambda$-
fluctuations on $\chi$. As in the case of gauge fluctuations, we need the
polarization operator $\Pi_{\lambda}(q, \omega)$ for $\lambda$-field, which
is the dynamical susceptibility near ${\bf q}_{AFM}=(\pi,\pi)$, (Fig.2). To be
consistent with previous consideration, only static part of $\Pi_{\lambda}(q,
\omega)$ has to be kept. The same arguments about long-distance behaviour
permit us to use an expansion over $q^2/m^2$, and we obtain [\onlinecite{yale}]
$$\Pi_{\lambda}(q,0)=-\frac{1}{\beta}\sum_{l}\int\frac{d^2 {\bf p}}{(2\pi)^2}
G(p,\epsilon_l)G(p+q,\epsilon_l)(i)^2=
\frac{1}{\pi \beta (q^2 + 4m^2)} \eqno(14)$$
Correspondingly, the $\lambda$-field temperature Green's function is
$D^{\lambda}(q)=-\frac{1}{N \Pi_{\lambda}(q,0)}$.
Unlike the gauge fluctuations, the fluctuations of the constraint field do not
produce infrared divergency in the $z$-particle self-energy.
Nevertheless, they give correction to $\chi$ of the same order of magnitute as
${\bf A}$-fluctuations. This can be shown by the calculation of the diagrams of
Fig.3
, using instead of $D_{\mu \nu}$ the $\lambda$-field Green's function $D_{
\lambda}$ and the corresponding vertex $i$. By this way we obtained (again
using
$m$ as an upper cut-off in the ${\bf q}$-integration)
$$\chi_{\lambda}^{1/N}=\frac{0.875}{4\pi \beta m^2} \eqno(15)$$
Thus, being relatively unimportant for single-particle properties, fluctuations
of the constraint field are as significant for the spin susceptibility as those
of the gauge field. Summing both $1/N$ corrections together, finally we have
$$\chi^{mf}(q=0,\omega=0)+\chi^{1/N}(q=0,\omega=0)=
\chi^{mf}(q=0,\omega=0)(1+\frac{1.575}{N}) \eqno(16)$$

The resulting effect of the fluctuations is to enhance the short range
antiferromagnetic order, in agreement with the results of [\onlinecite{yale}].
Interestingly, this enhancement arises from the integration over spatial scales
which are
 much longer than the correlation length. At the same time, the renormalization
of
the correlation length and the damping effects arise at smaller spatial scale,
as was shown in [\onlinecite{yale}].

Equation $(16)$ shows that both types of fluctuations are important in the
problem under consideration. To obtain the complete $1/N$ correction to the
susceptibility it is necessary to carry out the calculations in the entire
energy and momentum region, which is much more complicated problem
[\onlinecite{yale}]
and is beyond the scope of our paper.

In conclusion, we have shown that the spin-$1/2$ mean-field excitations
in the thermally disordered phase of the quantum antiferromagnet are
suppresed by the interaction with generated gauge massless field. At the same
time this interaction enhances the susceptibility, which describes the spin-$1$
collective excitations. Analogous enhancement is produced by the massive
fluctuations of the constraint field.

\acknowledgements

The authors would like to thank Z.Y.Weng for collaboration on the early stage
of this work and helpful discussions, and D.N.Sheng and C.S.Ting for useful
remarks. We'd like to acknowledge also useful comments of the referee.
 One of us (O.A.S.) is grateful to
 A.F.Barabanov, A.G.Abanov, V.Brazhnikov, and I.Gruzberg for
stimulating discussions. The present work
is supported by a grant from Robert A.Welch Foundation and the Texas Center
for Superconductivity at the University of Houston.

\begin{table}
\caption{Numerical values of $C_i$, $S_i$, and symmetry factors $F$ of
diagrams from Fig.3}\label{values}
\begin{tabular}{cccc}
\multicolumn{1}{c}{$Diagram$} &\multicolumn{1}{c}{$C$} &\multicolumn{1}{c}{$S$}
&\multicolumn{1}{c}{$F$}\\
\tableline
$a$ &1/6 &-1/20 &2 \\
$b$ &-1/4 &0 &2 \\
$c$ &1/6 &-1/60 &1 \\
\end{tabular}
\end{table}
\end{document}